\documentclass[conference]{IEEEtran}
\IEEEoverridecommandlockouts

\usepackage[T1]{fontenc}
\usepackage[utf8]{inputenc}
\usepackage[USenglish]{babel}
\usepackage[autostyle=once,english=american]{csquotes}

\usepackage{cite}
\usepackage{amsmath}
\usepackage{amssymb,amsfonts}
\usepackage{algorithm}
\usepackage{algorithmicx}
\usepackage{algpseudocode}
\usepackage{graphicx}
\usepackage{textcomp}
\usepackage{xcolor}
\usepackage{booktabs}
\usepackage{flushend}
\usepackage{url}
\usepackage{amsmath}
\usepackage{dblfloatfix}
\usepackage{balance}

\usepackage{tabularx,booktabs}
\usepackage{multirow}
\newcolumntype{C}{>{\centering\arraybackslash}X} 
\usepackage{amssymb}
\usepackage{pifont}
\newcommand{\cmark}{\ding{51}}%
\newcommand{\xmark}{\ding{55}}%

\usepackage{lipsum}
\usepackage{graphicx}

\makeatletter
\newcommand*{\defeq}{\mathrel{\rlap{%
			\raisebox{0.3ex}{$\m@th\cdot$}}%
		\raisebox{-0.3ex}{$\m@th\cdot$}}%
	=}
\makeatother

\def\BibTeX{{\rm B\kern-.05em{\sc i\kern-.025em b}\kern-.08em
    T\kern-.1667em\lower.7ex\hbox{E}\kern-.125emX}}

\begin{document}

\title{Decentralized Finance, Centralized Ownership? \\ An Iterative Mapping Process to Measure \\ Protocol Token Distribution}

\author{
\IEEEauthorblockN{Matthias Nadler}
    \IEEEauthorblockA{
    \textit{Center for Innovative Finance} \\
    \textit{Faculty of Business and Economics} \\
    \textit{University of Basel}\\
        Basel, Switzerland \\
        matthias.nadler@unibas.ch
    }
    \and
\IEEEauthorblockN{Fabian Schär}
    \IEEEauthorblockA{
    \textit{Center for Innovative Finance} \\
    \textit{Faculty of Business and Economics} \\
    \textit{University of Basel}\\
        Basel, Switzerland \\
        f.schaer@unibas.ch
    }
}
 
\maketitle

\begin{abstract}
In this paper, we analyze various Decentralized Finance (DeFi) protocols in terms of their token distributions. We propose an iterative mapping process that allows us to split aggregate token holdings from custodial and escrow contracts and assign them to their economic beneficiaries. This method accounts for liquidity-, lending-, and staking-pools, as well as token wrappers, and can be used to break down token holdings, even for high nesting levels. We compute individual address balances for several snapshots and analyze intertemporal distribution changes. In addition, we study reallocation and protocol usage data, and propose a proxy for measuring token dependencies and ecosystem integration. The paper offers new insights on DeFi interoperability as well as token ownership distribution and may serve as a foundation for further research.
\end{abstract}


\section{Introduction}

Decentralized Finance (DeFi) refers to a composable and trust-minimized protocol stack that is built on public Blockchain networks and uses smart contracts to create a large variety of publicly accessible and interoperable financial services. In contrast to traditional financial infrastructure, these services are mostly non-custodial and can mitigate counterparty risk without the need for a centralized third party. Funds are locked in smart contracts and handled in accordance with predefined rules, as specified by the contract code. 
Some examples of DeFi protocols include constant function market makers, lending-platforms, prediction markets, on-chain investment funds, and synthetic assets, \cite{schar2020defi}.

Most of these protocols issue corresponding tokens that represent some form of partial protocol ownership. Although the exact implementations, the feature sets, and the token holder rights vary greatly among these tokens, the reason for their existence can usually be traced back to two motives: \emph{Protocol Governance} and \emph{Protocol Economics}.

\begin{enumerate}
	\item[] \textbf{Governance:} Tokens may entitle the holder to vote on contract upgrades or parameter changes. A token-based governance system allows for the implementation of new features. Moreover, the protocol can react to exogenous developments, upcoming interface changes, and potential bugs. \vspace{0.5em}
	\item[] \textbf{Economics:} Most tokens have some form of implicit or explicit value-capture that allows the token holder to participate economically in the growth of the protocol. Value is usually distributed through a utility and burn mechanism (deflationary pressure) or some form of dividend-like payments. In many cases, initial token sales are used to fund protocol development and continuous release schedules to incentivize protocol usage.
\end{enumerate} \vspace{0.5em}

Considering the two main reasons for the existence of these tokens, it becomes apparent that token distribution is a critical factor in the protocols' decentralization efforts. Heavily centralized token allocations may result in situations where a small set of super-users can unilaterally change the protocol -- potentially at the expense of everyone else. Moreover, a heavily concentrated distribution may create an ecosystem where much of the value is captured by a small number of actors.

The authors are unaware of previous academic research on this subject. In August 2020, an analysis was circulated on social media, \cite{conti2020}. Simone Conti analyzed token contracts for their top holders and used this data to compute ownership concentration measures. However, the study was based on questionable assumptions and fails to account for the large variety of contract accounts. In particular, liquidity-, lending- and staking-pools, as well as token wrappers, had been counted as individual entities. As these contract accounts are mere custodians and usually hold significant token amounts on behalf of a large set of economic agents, this approach clearly leads to spurious results.

There are previous studies that tackle similar research questions in the context of the Bitcoin network, \cite{gupta2017gini}, \cite{chohan2019cryptocurrencies}, \cite{kondor2014rich}. However, due to Bitcoin's relatively static nature and the separation of token ownership and protocol voting rights, the question is less pressing. Moreover, the fact that Bitcoin's standard client discourages address reuse makes these analyses much harder to perform. In a similar vein, a recent working paper conducted an analysis for the evolution of shares in proof-of-stake based cryptocurrencies, \cite{rosu2020evolution}.

The remainder of this paper is structured as follows: In Section \ref{sec:sampleSelection} we describe how the token and snapshot samples have been selected. Sections \ref{sec:dataPreparation} and \ref{sec:dataAnalysis} explore the data preparation and analysis respectively. In Section \ref{sec:discussion} we discuss the results, limitations and further research. In Section \ref{sec:conclusion} we briefly summarize our findings and the contribution of this paper.

\section{Sample Selection}
\label{sec:sampleSelection}

In this section, we describe the scope of our analysis. In particular, we discuss how tokens and snapshots have been selected. 
The token selection determines which assets we observe. The snapshot selection determines at which point in time the blockchain state is observed. 

\subsection{Token Selection}

To qualify for selection, tokens had to fulfill the following criteria:

\begin{enumerate}
	\item The token must be a protocol token. It must incorporate some form of governance and/or utility mechanism. Pure stablecoins, token wrappers, or token baskets have not been considered.\footnote{Although wrappers and baskets will be considered for fund reallocation, as described in Section \ref{sec:dataPreparation}.}
	\item The token must be \texttt{ERC-20} compliant and contribute towards decentralized financial infrastructure.
	\item As of September 15th, 2020, the token must fulfill at least one of the following three conditions:
	\begin{itemize}
		\item[a)] Relevant supply with market cap $\geq$ 200 mm (MC).
		\item[b)] Total value locked in the protocol's contracts (vesting not included) $\geq$ 300 mm (VL).
		\item[c)] Inclusion in Simone Conti's table (SC). 	
	\end{itemize}
\end{enumerate}

Market cap and value locked serve as objective and quantitative inclusion criteria. Tokens from Simone Conti's table have mainly been included to allow for comparisons. 

Applying these criteria, we get a sample of 18 DeFi tokens. The tokens and the reason for their selection are summarized in Table \ref{tbl:tokens}. Please note that we have decided to exclude SNX since some of its features are not in line with standard conventions and make it particularly difficult to analyze.

\begin{table}[h!]
	\caption{Token Selection}
	\label{tbl:tokens}
\center
	\begin{tabular}{lcccr}
	\toprule
		Token & MC     & VL     & SC     & Deployment  \\  \midrule
		BAL   & \xmark & \cmark & \cmark & 2020-06-20  \\
		BNT   & \xmark & \xmark & \cmark & 2017-06-10  \\
		COMP  & \cmark & \cmark & \cmark & 2020-03-04  \\
		CREAM & \xmark & \cmark & \xmark & 2020-08-04  \\
		CRV   & \xmark & \cmark & \xmark & 2020-08-13  \\
		KNC   & \cmark & NA     & \cmark & 2017-09-12  \\
		LEND  & \cmark & \cmark & \cmark & 2017-09-17  \\ 
		LINK  & \cmark & NA     & \xmark & 2017-09-16  \\
		LRC   & \cmark & \xmark & \xmark & 2019-04-11  \\ 
		MKR   & \cmark & \cmark & \cmark & 2017-11-25  \\
		MTA   & \xmark & \xmark & \cmark & 2020-07-13  \\
		NXM   & \cmark & \xmark & \xmark & 2019-05-23  \\
		REN   & \cmark & \xmark & \cmark & 2017-12-31  \\
		SUSHI & \cmark & \cmark & \xmark & 2020-08-26  \\
		UMA   & \cmark & \xmark & \xmark & 2020-01-09  \\  
		YFI   & \cmark & \cmark & \cmark & 2020-07-17  \\ 
		YFII  & \cmark & \xmark & \xmark & 2020-07-26  \\
		ZRX   & \cmark & NA     & \xmark & 2017-08-11  \\  \bottomrule
	\end{tabular}
\end{table}

\subsection{Snapshot Selection}
To analyze how the allocation metrics change over time, we decided to conduct the analysis for various snapshots. The first snapshot is from June 15th, 2019. We had then taken monthly snapshots. The snapshots' block heights and timestamps are listed in Table \ref{tbl:snapshots}.

\begin{table}[h!]
	\caption{Snapshot Selection}
	\label{tbl:snapshots}
\center
	\begin{tabular}{lll}
	\toprule
		Nr. & Block Height   & Date \\ \midrule
		1 & 7962629  & 2019-06-15   \\
		2 & 8155117  & 2019-07-15   \\
		3 & 8354625  & 2019-08-15   \\
		4 & 8553607  & 2019-09-15   \\
		5 & 8745378  & 2019-10-15   \\
		6 & 8938208  & 2019-11-15   \\
		7 & 9110216  & 2019-12-15   \\
        8 & 9285458  & 2020-01-15   \\
        9 & 9487426  & 2020-02-15   \\
        10& 9676110  & 2020-03-15   \\
        11& 9877036  & 2020-04-15   \\
        12& 10070789 & 2020-05-15   \\
        13& 10270349 & 2020-06-15   \\
        14& 10467362 & 2020-07-15   \\
        15& 10664157 & 2020-08-15   \\
        16& 10866666 & 2020-09-15   \\ \bottomrule
	\end{tabular}
\end{table}

\section{Data Preparation}
\label{sec:dataPreparation}

We use our token and snapshot selection from \ref{sec:sampleSelection} to analyze the allocation characteristics and observe how they change over time. All the necessary transaction- and event data was directly extracted from a Go-Ethereum node using Ethereum-ETL, \cite{ethereumetl}. To construct accurate snapshots of token ownership, we must map each token holding to the address that actually owns and may ultimately claim the funds. 

A simple example is the YFI/wETH Uniswap V2 liquidity pool: A naïve analysis would lead to the conclusion that the tokens are owned by the Uniswap exchange contract. However, this contract is just a liquidity pool with very limited control over the tokens it holds. Full control, and thus ownership of the tokens, remains with the liquidity providers. To account for this and to correctly reflect the state of token ownership, the tokens must be mapped proportionally from the exchange contract to the liquidity providers.

A more complex example illustrates the need for an iterative mapping process: YFI is deposited into a Cream lending pool, minting crYFI for the owner. This crYFI together with crCREAM is then deposited in a crYFI/crCREAM Balancer-like liquidity pool, minting CRPT (Cream pool tokens) for the depositor. Finally, these CRPT are staked in a Cream staking pool, which periodically rewards the staker with CREAM tokens but does not mint any ownership tokens. The actual YFI tokens, in this case, are held by the Cream lending pool. Trying to map them to their owners via the lending pool tokens (crYFI) will lead us to the liquidity pool and finally to the staking pool, where we can map the YFI to the accounts that staked the CRPT tokens. Each of these steps needs to be approached differently, as the underlying contracts have distinct forms of tracking token ownership. And further, these steps must also be performed in the correct order.

\subsection{Identifying and Categorizing Addresses}
\label{sub:identifyCategorize}
Addresses that do not have bytecode deployed on them - also called externally owned accounts or EOAs - cannot be analyzed further with on-chain data. To determine whether to include or exclude an EOA from our analysis, we use a combination of tags from etherscan.io, nansen.ai, and coingecko.com, \cite{etherscan}, \cite{nansen}, \cite{coingecko}. An EOA qualifies for exclusion if it is a known burner address, owned by a centralized, off-chain exchange (CEX) or if the tokens on the account are disclosed by the developer team as FTIA (foundation, team, investor, and advisor) vesting. Every other EOA is assumed to be a single actor and is included in the analysis. 

Addresses with deployed bytecode are smart contracts or contract accounts. These contracts are analyzed and categorized based on their ABI, bytecode, return values, and manual code review. Most implementations of multisig wallets are detected and treated equivalent to EOAs. Mappable smart contracts are described by the following categories:
\begin{enumerate}
	\item[] \textbf{Liquidity Pools:} Decentralized exchanges, converters, token baskets, or similar contracts that implement one or more \texttt{ERC-20} liquidity pool tokens. The funds are mapped proportionally to the relevant liquidity pool tokens. \vspace{0.5em}
	\item[] \textbf{Lending Pools:} 	Aave, Compound, and Cream offer lending and borrowing of tokens. Both the debts and deposits are mapped to their owners using protocol-specific events and archival calls to the contracts. \vspace{0.5em}
	\item[] \textbf{Staking Contracts:} Staking contracts differ from liquidity pools in the sense that they usually do not implement an \texttt{ERC-20} token to track the stakes of the owners. We further differentiate if the token in question is used as a reward, as a stake, or both. Future staking rewards are excluded as they cannot be reliably mapped to future owners. The remaining tokens are mapped using contract-specific events for depositing and withdrawing stakes and rewards. For Sushi-like staking pools, we also account for a possible migration of staked liquidity pool tokens. \vspace{0.5em}
	\item[] \textbf{Unique Contracts:} These contracts do not fit any of the above categories, but the tokens can still be mapped to their owners. Each contract is treated individually, using contract-specific events and archival calls where needed. A few examples include MKR governance voting, REN darknode staking, or LRC long-term holdings.
\end{enumerate} \vspace{0.5em}

Smart contracts which hold funds that are not owned by individual actors or where no on-chain mapping exists are excluded from the analysis. Most commonly, this applies to contracts that hold and manage funds directly owned by a protocol with no obvious distribution mechanism.

\subsection{Iterative Mapping Process for Tokens}
For each token and snapshot, we construct a token holder table listing the initial token endowments per address. We then proceed with an iterative mapping process as follows:

\begin{algorithm}[H]
	\caption{Iterative Mapping Process}
	\begin{algorithmic}[1]
		\State $H \gets$ initial token holder table
		\Repeat 
			\State sort $H$ by token value, descending
			\ForAll{$h \in$ top 1,000 rows of $H$}
			\State identify and categorize $h$
			\State apply inclusion logic to $h$
			\If{$h$ is mappable}
			\State map $h$ according to its category
			\EndIf
			\EndFor
		\Until{no mappable rows found in last iteration}
		\State \textbf{assert} every row with more than 0.1\% of the total relevant supply is properly identified and categorized
	\end{algorithmic}
\end{algorithm}

It is possible that tokens must be mapped from an address onto themselves. For most mappable contracts, these tokens are permanently lost\footnote{For example, if Uniswap liquidity pool tokens are directly sent to their liquidity pool address, they can never be retrieved.} and are thus treated as burned and are excluded from the analysis. For contracts where the tokens are not lost in this way, we implemented contract-specific solutions to avoid potential infinite recursion.

Every instance of a remapping from one address to another, called an adjustment, is tracked and assigned to one of five adjustment categories. There is no distinction between situations where the protocol token or a wrapped version thereof is remapped. The five adjustment categories are:

\begin{enumerate}
	\item[] \textbf{Internal Staking:} Depositing the token into a contract that is part of the same protocol. This includes liquidity provision incentives, protocol stability staking, and some forms of governance voting.\vspace{0.5em}
	\item[] \textbf{External Staking:} Depositing the token into a contract that is not part of the same protocol. This is most prominent for Sushi-like liquidity pool token staking with the intention of migrating the liquidity pool tokens, but it also includes a variety of other, external incentive programs.\vspace{0.5em}
	\item[] \textbf{AMM Liquidity:} Depositing the token into a liquidity pool run by a decentralized exchange with some form of an automated market maker.\vspace{0.5em}
	\item[] \textbf{Lending / Borrowing:} Depositing the token into a liquidity pool run by a decentralized lending platform or borrowing tokens from such a pool. \vspace{0.5em}
	\item[] \textbf{Other:} Derivatives, 1:1 token wrappers with no added functionality, token migrations, and investment fund-like token baskets.
\end{enumerate}  \vspace{0.5em}

\begin{table*}
    \caption{Token Ownership Structure}
    \label{tbl:holdings}
    \begin{tabularx}{\textwidth}{@{}lr*{8}{C}c@{}}
        \toprule
        Token & & Owner \# & Top 5 & Top 10 & Top 50 & Top 100 & Top 500 & Top 50\%   & Top 99\% & Gini 500  \\
        \midrule
        BAL$\dagger$ & Sep 20 & 16,661 & 27.6\% & 36.71\% & 77.3\% & 85.01\% & 94.86\% & 18 & 2,157 & 83.77\% \\
        \midrule
        \multirow{3}{*}{BNT} & Sep 20 & 49,294 & 15.69\% & 24.71\% & 49.5\% & 61.77\% & 80.95\% & 52 & 10,010 & 69.82\% \\
        & Trend & +1.64\% & -5.43\% & -4.43\% & -2.94\% & -2.14\% & -1.06\% & +49.45\% & +7.52\% & -1.5\% \\
        & $\sigma$ 12m & 2,882.0 & 0.0712 & 0.0764 & 0.0827 & 0.0669 & 0.0378 & 15.7 & 1,481.9 & 0.0487 \\
        \midrule
        COMP$\dagger$ & Sep 20 & 36,033 & 31.23\% & 43.79\% & 86.75\% & 96.15\% & 98.91\% & 14 & 564 & 90.36\% \\
        \midrule
        CREAM$\dagger$ & Sep 20 & 4,426 & 48.44\% & 57.11\% & 74.32\% & 81.77\% & 94.16\% & 6 & 1,549 & 83.04\% \\
        \midrule
        CRV$\dagger$ & Sep 20 & 11,076 & 56.92\% & 61.09\% & 73.23\% & 79.07\% & 90.27\% & 2 & 3,549 & 84.64\% \\
        \midrule
        \multirow{3}{*}{KNC} & Sep 20 & 92,780 & 24.93\% & 35.63\% & 57.73\% & 64.62\% & 77.99\% & 26 & 19,922 & 77.6\% \\
        & Trend & +6.51\% & +3.36\% & +5.01\% & +2.14\% & +0.98\% & +0.04\% & -5.39\% & +15.74\% & +1.21\% \\
        & $\sigma$ 12m & 12,589.4 & 0.0302 & 0.0594 & 0.0489 & 0.0336 & 0.0171 & 13.9 & 3,971.3 & 0.0374 \\
        \midrule
        \multirow{3}{*}{LEND} & Sep 20 & 174,861 & 36.67\% & 43.64\% & 61.44\% & 67.42\% & 80.05\% & 16 & 57,534 & 79.97\% \\
        & Trend & +0.23\% & +33.26\% & +22.23\% & +11.35\% & +8.26\% & +3.74\% & -9.77\% & -4.7\% & +3.98\% \\
        & $\sigma$ 12m & 3,066.9 & 0.1294 & 0.1389 & 0.1358 & 0.1258 & 0.0878 & 82.2 & 21,962.9 & 0.0933 \\
        \midrule
        \multirow{3}{*}{LINK} & Sep 20 & 233,128 & 7.18\% & 13.46\% & 37.0\% & 44.99\% & 61.23\% & 166 & 61,910 & 65.27\% \\
        & Trend & +31.34\% & -0.5\% & -0.62\% & +1.72\% & +1.24\% & +0.08\% & -2.73\% & +16.99\% & +1.24\% \\
        & $\sigma$ 12m & 52,004.9 & 0.0029 & 0.004 & 0.0221 & 0.0204 & 0.0067 & 25.0 & 12,158.7 & 0.0279 \\
        \midrule
        \multirow{3}{*}{LRC} & Sep 20 & 66,382 & 13.75\% & 20.06\% & 43.44\% & 62.11\% & 87.9\% & 66 & 5,251 & 66.36\% \\
        & Trend & +1.49\% & -2.3\% & -1.68\% & -1.26\% & -1.14\% & -0.41\% & +3.23\% & +7.95\% & -0.74\% \\
        & $\sigma$ 12m & 3,392.5 & 0.0236 & 0.0232 & 0.0261 & 0.0313 & 0.0163 & 6.1 & 811.7 & 0.0205 \\
        \midrule
        \multirow{3}{*}{MKR} & Sep 20 & 29,765 & 24.43\% & 36.49\% & 67.71\% & 79.49\% & 93.72\% & 20 & 3,918 & 79.26\% \\
        & Trend & +8.31\% & -3.45\% & -2.12\% & -0.45\% & -0.19\% & -0.12\% & +4.5\% & +7.17\% & -0.22\% \\
        & $\sigma$ 12m & 4,511.7 & 0.0503 & 0.0405 & 0.0175 & 0.0107 & 0.0057 & 3.0 & 587.0 & 0.01 \\
        \midrule
        MTA$\dagger$ & Sep 20 & 5,595 & 13.81\% & 22.97\% & 51.18\% & 63.51\% & 88.27\% & 47 & 2,090 & 65.93\% \\
        \midrule
        \multirow{3}{*}{NXM} & Sep 20 & 7,355 & 32.17\% & 44.3\% & 70.42\% & 78.51\% & 91.29\% & 14 & 2,817 & 81.14\% \\
        & Trend & -36.69\% & -2.87\% & -2.71\% & -1.65\% & -1.12\% & -0.37\% & +18.09\% & -33.11\% & -0.24\% \\
        & $\sigma$ 12m & 1,918.2 & 0.0704 & 0.0992 & 0.0869 & 0.0619 & 0.0238 & 2.7 & 747.1 & 0.0434 \\
        \midrule
        \multirow{3}{*}{REN} & Sep 20 & 22,770 & 10.45\% & 15.29\% & 32.81\% & 41.79\% & 67.85\% & 166 & 8,500 & 55.31\% \\
        & Trend & +26.0\% & -3.12\% & -2.97\% & -2.98\% & -2.64\% & -1.5\% & +42.78\% & +25.39\% & -1.56\% \\
        & $\sigma$ 12m & 4,673.4 & 0.0232 & 0.0313 & 0.0671 & 0.072 & 0.0579 & 38.4 & 1,718.0 & 0.0437 \\
        \midrule
        SUSHI$\dagger$ & Sep 20 & 22,740 & 25.64\% & 35.26\% & 58.31\% & 66.28\% & 83.78\% & 28 & 7,300 & 74.11\% \\
        \midrule
        UMA$\dagger$ & Sep 20 & 5,634 & 56.21\% & 75.64\% & 96.87\% & 98.21\% & 99.43\% & 5 & 240 & 95.61\% \\
        \midrule
        YFI$\dagger$ & Sep 20 & 14,296 & 11.52\% & 16.98\% & 37.32\% & 48.1\% & 73.75\% & 114 & 5,145 & 57.6\% \\
        \midrule
        YFII$\dagger$ & Sep 20 & 8,513 & 20.8\% & 27.78\% & 53.93\% & 66.23\% & 85.15\% & 40 & 3,278 & 72.18\% \\
        \midrule
        \multirow{3}{*}{ZRX} & Sep 20 & 161,285 & 23.71\% & 38.4\% & 59.39\% & 63.87\% & 72.91\% & 21 & 38,404 & 82.63\% \\
        & Trend & +4.05\% & -1.15\% & -0.02\% & +0.76\% & +0.64\% & +0.22\% & -2.96\% & +6.28\% & +0.43\% \\
        & $\sigma$ 12m & 16,372.0 & 0.0133 & 0.0056 & 0.0158 & 0.0147 & 0.0082 & 3.6 & 5,233.6 & 0.0132 \\
        \midrule
        \multicolumn{10}{l}{$\dagger$\footnotesize{Insufficient historical data.}} \\
        \bottomrule
    \end{tabularx}
\end{table*}

\section{Data Analysis}
\label{sec:dataAnalysis}

In this section, we will use our data set to analyze two questions: \emph{First}, we study the token ownership concentration and use our remapping approach to compute more accurate ownership tables and introduce new allocation metrics. These metrics are of particular interest, as highly concentrated token allocations could potentially undermine any decentralization efforts. \emph{Second}, we use our remapping and protocol usage data to introduce wrapping complexity, shortage, and token interaction measures. These measures essentially serve as a proxy and indicate the degree of integration into the DeFi ecosystem. Moreover, they may serve as an important measure for potential dependencies and the general stability of the system. 

\subsection{Concentration of Token Ownership}

Table \ref{tbl:holdings} shows key metrics to illustrate the concentration of adjusted token ownership for the most recent snapshot, September 15th, 2020. The table is described below. Please note that \emph{relevant supply} refers to the sum of all adjusted and included token holdings, taking into account outstanding debts. Excluded token holdings are described in detail in section \ref{sub:identifyCategorize}.

\begin{enumerate}
	\item[] \textbf{Owner \#:} Total number of addresses owning a positive amount or fraction of the token. \vspace{0.5em}
	\item[] \textbf{Top n:} Percentage of the relevant supply held by the top $n$ addresses. \vspace{0.5em}
	\item[] \textbf{Top n\%:} Minimum number of addresses owning a combined $n\%$ of the relevant supply. \vspace{0.5em}
	\item[] \textbf{Gini 500:} The Gini coefficient, \cite{gini1912variabilita}, is used to show the wealth distribution inequality among the top 500 holders of each token. It can be formalized as \eqref{eq:gini}.
	\begin{equation}
	\label{eq:gini}
	G_{500} = \frac{\sum_{i=1}^{500}\sum_{j=1}^{500}\lvert x_i-x_j \rvert}{2 \cdot 500 ^2\bar{x}}
	\end{equation}
\end{enumerate} \vspace{0.5em}

For tokens with historical data of at least 12 months, we include the trend and standard deviation over this period. The trend represents the monthly change in percent according to an OLS regression line; the standard deviation shows the volatility of the trend.

\subsection{Ecosystem Integration}

Table \ref{tbl:wrap} presents key metrics of the tokens' integration into the DeFi ecosystem. The table is described below. 

\begin{enumerate}
	\item[] \textbf{Inclusion \%:} Relevant token supply divided by total token supply, excluding burned tokens. \vspace{0.5em}
	\item[] \textbf{Wrapping Complexity:} Relevant adjustments divided by relevant supply. This includes only adjustments to non-excluded addresses\footnote{Some of the excluded addresses still deposit their tokens in mappable contracts; e.g. a CEX depositing their users' tokens in a staking pool. To prevent distortion, we exclude these funds from both the relevant supply and the relevant adjustments.} and may (in extreme cases) reach values above 1. The Wrapping complexity is formalized in \eqref{eq:wrap}, where $\pmb{\omega}:=(\omega_1,\dots, \omega_N)$ represents the vector of all relevant adjustments for a given token and $\bar{S}$ represents relevant supply. \vspace{0.5em}
	\begin{equation}
	\label{eq:wrap}
	\frac{\sum_{i = 1}^{N} \left|\omega_{i} \right| }{\bar{S}}
\end{equation}
	\item[] \textbf{Multi-Token Holdings:} Number of addresses with a minimum allocation of 0.1\% of this token and 0.1\% for at least $n\in(1,2,3,4)$ other tokens from our sample. \vspace{0.5em}
	\item[] \textbf{Shorted:} Negative token balances in relation to relevant supply; i.e. value on addresses that used lending markets to borrow and resell the token, to obtain a short exposure, divided by $\bar{S}$.
\end{enumerate} \vspace{0.5em}

It is important to note that the inclusion ratio is predominantly dictated by the tokens' emission schemes. In some cases, the total supply is created with the \texttt{ERC-20} token deployment but held in escrow and only released over the following years. Consequently, we excluded this non-circulating supply.

\begin{table*}
    \caption{Token Wrapping Complexity}
    \label{tbl:wrap}
    \begin{tabularx}{\textwidth}{llccccccccccccccc}
        \toprule
        \multirow{2}{*}{Token} & {} & \multirow{2}{*}{Inclusion \%} & {} &
        \multicolumn{6}{c}{Wrapping Complexity} & {} & \multicolumn{4}{c}{Multi-Token Holdings} & {} &
        \multirow{2}{*}{Shorted} \\
        & & & & Jun-19 & Sep-19 & Dec-19 & Mar-20 & Jun-20 & Sep-20 & & 1+ & 2+ & 3+ & 4+  \\
        \cline{1-1} \cline{3-3} \cline{5-10} \cline{12-15} \cline{17-17} \\

        BAL & {} & 19.6\% & {} & - & - & - & - & - & 51.7\%& {} & 17.6\% & 5.5\% & 1.1\% & -& {} & 0.026\% \\
        BNT & {} & 56.8\% & {} & 11.9\% & 11.9\% & 10.3\% & 20.8\% & 9.6\% & 10.2\%& {} & 8.7\% & 1.4\% & 0.7\% & 0.7\%& {} & - \\
        COMP & {} & 36.0\% & {} & - & - & - & 0.0\% & 0.0\% & 7.5\%& {} & 8.4\% & 3.6\% & 2.4\% & -& {} & 0.004\% \\
        CREAM & {} & 3.6\% & {} & - & - & - & - & - & 455.0\%& {} & 30.1\% & 11.8\% & 5.4\% & -& {} & 11.971\% \\
        CRV & {} & 2.2\% & {} & - & - & - & - & - & 43.1\%& {} & 20.9\% & 9.9\% & 4.4\% & 2.2\%& {} & 0.761\% \\
        KNC & {} & 70.7\% & {} & 0.2\% & 0.2\% & 0.4\% & 2.9\% & 1.8\% & 48.4\%& {} & 17.7\% & 9.4\% & 4.2\% & 2.1\%& {} & 0.123\% \\
        LEND & {} & 69.3\% & {} & 0.0\% & 0.0\% & 0.1\% & 28.9\% & 50.7\% & 63.1\%& {} & 38.6\% & 19.3\% & 6.8\% & 2.3\%& {} & 0.039\% \\
        LINK & {} & 31.3\% & {} & 0.0\% & 0.0\% & 0.0\% & 1.8\% & 2.2\% & 13.6\%& {} & 12.9\% & 5.9\% & 4.0\% & 2.0\%& {} & 0.383\% \\
        LRC & {} & 58.8\% & {} & 5.3\% & 4.7\% & 7.4\% & 19.0\% & 21.4\% & 23.1\%& {} & 1.8\% & 0.6\% & - & -& {} & - \\
        MKR & {} & 81.5\% & {} & 33.6\% & 23.2\% & 31.5\% & 28.6\% & 37.3\% & 41.5\%& {} & 7.2\% & 2.4\% & 0.8\% & -& {} & 0.036\% \\
        MTA & {} & 3.1\% & {} & - & - & - & - & - & 73.8\%& {} & 15.1\% & 4.8\% & 1.8\% & -& {} & 2.631\% \\
        NXM & {} & 95.1\% & {} & 0.0\% & 0.0\% & 0.0\% & 0.0\% & 0.0\% & 66.7\%& {} & 17.0\% & 8.0\% & 2.0\% & -& {} & - \\
        REN & {} & 61.3\% & {} & 0.0\% & 0.0\% & 0.0\% & 0.2\% & 12.1\% & 59.9\%& {} & 11.4\% & 4.4\% & 3.2\% & 1.3\%& {} & 0.035\% \\
        SUSHI & {} & 48.2\% & {} & - & - & - & - & - & 109.9\%& {} & 28.9\% & 9.9\% & 1.7\% & -& {} & 0.844\% \\
        UMA & {} & 53.8\% & {} & - & - & - & 0.0\% & 0.4\% & 3.0\%& {} & 4.3\% & - & - & -& {} & - \\
        YFI & {} & 94.8\% & {} & - & - & - & - & - & 70.5\%& {} & 41.0\% & 14.1\% & 2.6\% & -& {} & 0.307\% \\
        YFII & {} & 40.1\% & {} & - & - & - & - & - & 54.2\%& {} & 8.6\% & 4.3\% & 1.4\% & -& {} & - \\
        ZRX & {} & 57.9\% & {} & 0.7\% & 1.9\% & 1.7\% & 4.5\% & 6.8\% & 32.8\%& {} & 19.0\% & 6.3\% & 4.8\% & 3.2\%& {} & 0.052\% \\
        \bottomrule
    \end{tabularx}
\end{table*}

\begin{figure*}
	\centering
	\caption{Adjustment Graphs}
	\label{fig:wrapTime}
  \includegraphics[width=\textwidth ,height=\textheight,keepaspectratio]{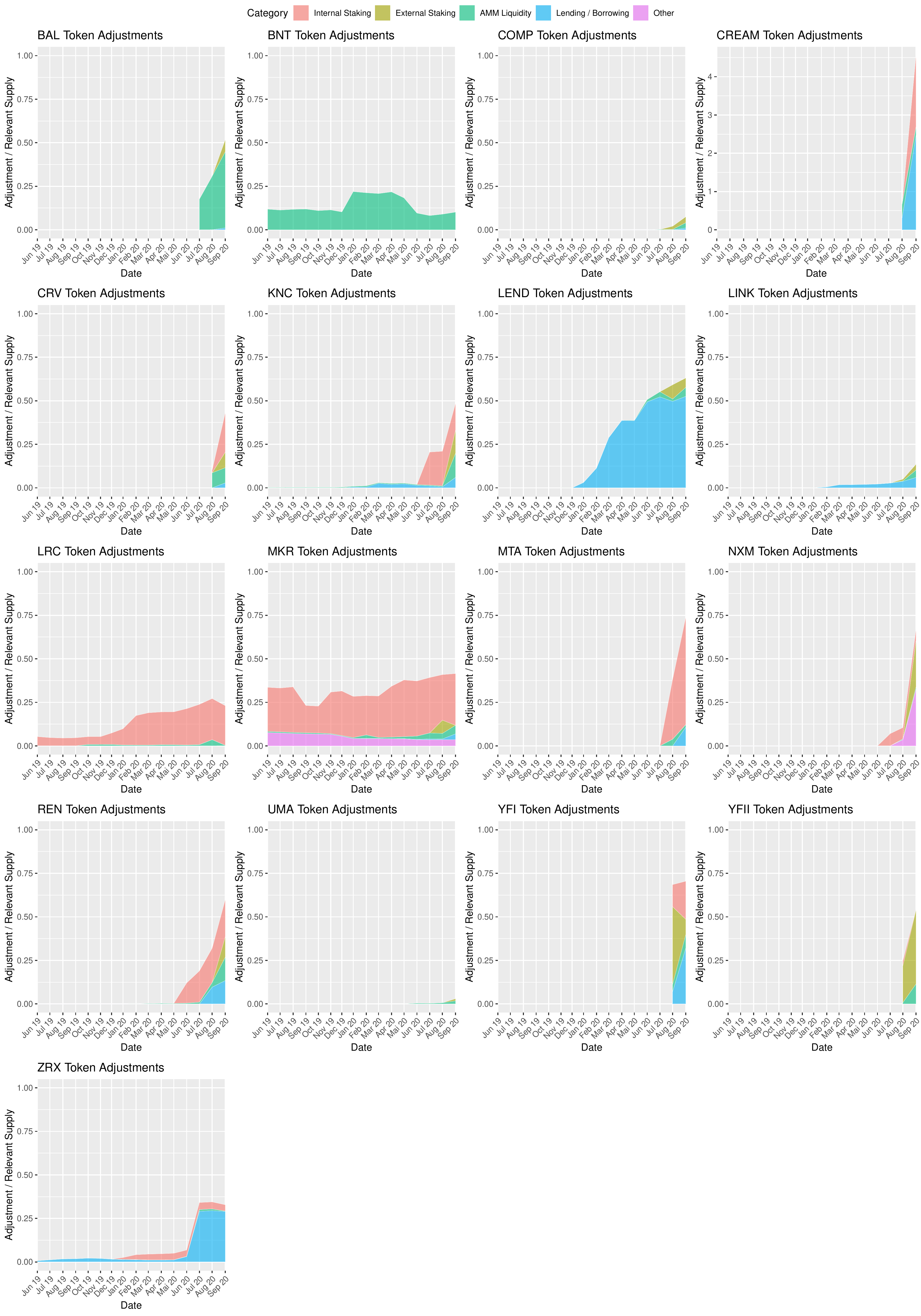}
\end{figure*}

Figure \ref{fig:wrapTime} shows the development of the tokens' wrapping complexities by adjustment category in a stacked time series. Note that the limits of the $y$-axis for the CREAM graph are adjusted to accommodate for the higher total wrapping complexity. We have not included a graph for the SUSHI token, as there is only one snapshot available since its launch\footnote{On September 15th, 2020, the 109.9\% wrapping complexity of SUSHI is composed of 28.2\% internal staking, 49.3\% external staking, 30.1\% AMM liquidity, and 2.2\% lending/borrowing.}.

A wrapping complexity $>1$ means that the same tokens are wrapped several times. If, for example, a token is added to a lending pool, borrowed by another person, subsequently added to an AMM liquidity pool, and the resulting LP tokens staked in a staking pool, the wrapping complexity would amount to 4. Similarly, a single token could be used multiple times in a lending pool and thereby significantly increase the wrapping complexity. 

Note that most tokens have experienced a sharp increase in wrapping complexity in mid-2020. The extent to which each category is used depends on the characteristics of each token; internal staking, in particular, can take very different forms.

The ``other'' category is mainly driven by token migrations, where new tokens are held in redemption contracts, and 1:1 token wrappers.

\vspace{1em}

\hspace{1em}

\section{Discussion}
\label{sec:discussion}

In this section we discuss the results from our data analysis. We revisit Table \ref{tbl:holdings} and \ref{tbl:wrap} as well as Figure \ref{fig:wrapTime} and discuss some interesting findings.


What seems to be true across the board is that DeFi tokens have a somewhat concentrated ownership structure. This is certainly an issue that merits monitoring, as it may potentially undermine many of the advantages this new financial infrastructure may provide.

For protocols with token-based governance models, the lower bound number of addresses needed to reach a majority, i.e., >50\%, may be of special interest. A relatively low threshold can indicate a higher likelihood of collusion and centralized decision making. In extreme cases, a few individuals could jointly enact protocol changes. However, since governance rules, the implementations of voting schemes, and security modules (e.g., timelocks) vary greatly between protocols, direct comparisons should only be made with great care. 

In addition to the decentralization and governance concerns, the study also shows DeFi's limitations with regard to transparency. While it is true that the DeFi space is extremely transparent in the sense that almost all data is available on-chain, it is very cumbersome to collect the data and prepare it in a digestible form. High nesting levels with multiple protocols and token wrappers involved will overwhelm most users and analysts and create the need for sophisticated analysis tools. The computation of accurate token ownership statistics and reliable dependency statistics is extremely challenging.

The problem becomes apparent when we compare our results to the results of Simone Conti's analysis, \cite{conti2020}. Recall that Conti's analysis has not controlled for any account-specific properties. Our analysis shows that for most tokens, the token holdings of the top 5 addresses thereby have been overestimated by approximately 100\% and in some extreme cases by up to 700\%. The main source of these errors is the inclusion of token holdings from custodial- and escrow contracts, such as liquidity-, lending-, and staking-pools, as well as token wrappers, vesting contracts, migrations, burner addresses, and decentralized exchange addresses. We control for these accounts and split their holdings to the actual beneficiary addresses where possible and exclude them where not possible. A closer comparison of the two tables reveals that the differences remain high for lower holder thresholds (i.e., top 10, top 50, and top 100). At the top 500 threshold, the differences are still significant, although to a much lesser degree. 


In addition to the computation of more accurate holder tables, transparency is a precondition for the analysis of protocol interconnections and dependencies. For this purpose, we introduce the wrapping complexity and multi-token holding metrics. Wrapping complexity essentially shows how the token is used in the ecosystem. On the one hand, high wrapping complexities can be interpreted as an indicator for a token that is deeply integrated into the DeFi ecosystem. 
On the other hand, high wrapping complexities may also be an indicator for convoluted and unnecessarily complex wrapping schemes that may introduce additional risks. 

A potential indicator for how the market feels about the complexity is the shortage percentage, i.e., the value of all decentralized short positions in relation to the relative supply. Interestingly, there is a high positive correlation between the two measures, which may at first glance suggest that wrapping complexity is interpreted as a negative signal. However, this would be a problematic interpretation since wrapping complexity is, in fact, at least partially driven by the shortage activity. Once we exclude the lending and borrowing, as well as ``other'' categories, the effect becomes less pronounced. 

The DeFi space is developing very rapidly and constantly increases in complexity. Many new and exciting protocols have emerged in 2020. Novel concepts such as complex staking schemes started to play a role in most protocols. We see staking, or more specifically staking rewards, as a catalyst for the immense growth in the DeFi space. However, it is somewhat questionable whether this growth will be sustainable. Treasury pools will eventually run out of tokens, and uncontrolled token growth leads to an increase of the relevant token supply, which may create inflationary pressure. 


While we are confident that our study provides interesting contributions with new metrics and processes to compute token ownership tables with unprecedented accuracy, we would still like to mention some of the limitations of our study and point out room for further extensions.

First, we perform no network analysis to potentially link multiple addresses of the same actor. This approach has likely lead to an overestimation of decentralization. In a further research project, one could combine our data set and remapping method with address clustering.

Second, while the automated process may remap tokens for all contract accounts, our manual analysis was limited to contract accounts with a significant amount. We decided to set the threshold value at 0.1\% of relevant supply.

Third, we used various data sources to verify the labeling of addresses. In some unclear cases, we approached the teams directly for more information. However, this information cannot be verified on-chain. Consequently, this is the only part of the study for which we had to rely on information provided by third parties.


Further research may adopt the methods of this paper to analyze token characteristics in the context of governance models. The data could be used as a parameter for more realistic simulations and game-theoretical governance models. Novel metrics, such as the wrapping complexity, may be useful for studies concerned with the interdependencies and risk assessment of the DeFi landscape. Finally, the proposed readjustment categories may provide a good base for further research on how DeFi tokens are being used and the reasons for their spectacular growth.

\section{Conclusion}
\label{sec:conclusion}
\balance
In this paper, we analyze the holder distribution and ecosystem integration for the most popular DeFi tokens. The paper introduces a novel method that allows us to split and iteratively reallocate contract account holdings over multiple wrapping levels. 

Our data indicate that previous analyses severely overestimated ownership concentration. However, in most cases, the majority of the tokens are still held by a handful of individuals. This finding may raise important questions regarding protocol decentralization and build a foundation for DeFi governance research. 

We further investigated dependencies and ecosystem integration. Our analysis suggests that the complexity of the ecosystem has drastically increased. This increase seems to be consistent among most tokens. However, the main drivers vary significantly, depending on the nature of the token. 

To conclude, DeFi is an exciting and rapidly growing new financial infrastructure. However, there is a particular risk that high ownership concentration and complex wrapping structures introduce governance risks, undermine transparency and create extreme interdependence affecting protocol robustness.

\bibliographystyle{IEEEtran}
\bibliography{ms_bib}

\section*{Acknowledgements}
The authors would like to thank Mitchell Goldberg, John Orthwein and Victoria J. Block for proof-reading the manuscript.

\end{document}